\documentclass[showpacs,floatfix,preprint,nofootinbib]{revtex4}
\usepackage{bm}
\let\om=\omega
\begin{document}
    \title{Testing MOND on Earth}
    \author{A.Yu.Ignatiev} 
    \address{Theoretical Physics Research Institute, Melbourne 3163, Australia} 
    \begin{abstract} 
    MOND is one of the most popular alternatives to Dark Matter (DM). While efforts to directly detect DM in laboratories have been steadily pursued over the years,  the proposed Earth-based tests of MOND are still in their infancy.  Some proposals recently appeared in the literature are briefly reviewed, and it is argued that collaborative efforts of theorists and experimenters are needed to move forward in this exciting new area. Possible future directions are outlined.
    \end{abstract}
    \maketitle 
\section{Introduction}
The existing astrophysical evidence suggests that Newton's second law may need a modification. This idea is part of the MOND paradigm \cite{m1}. The abbreviation stands for Modified Newtonian Dynamics.
Actually, there are 2 versions of MOND \cite{m1}: one is to modify the universal gravitational law, and the other is to change Newton's second  law. Here, we are only talking about the latter as the former has been extensively reviewed before (see Ref.\cite{FM} and references therein). 
  
This whole idea was discussed and tested  within astrophysical context. People looked into astronomy data. What was  suggested more recently, is that we should look at the laboratory experiments \cite{Ig1,Ig2}. We should look at how to test this idea in an ordinary experimental way that we test other similar ideas.

 The modification of dynamics takes effect only when the acceleration  is very small, of order $a_{0}\sim 10^{-10} m s^{-2}$.
This acceleration is much smaller than any acceleration  that we usually encounter as under ordinary conditions every physical body is acted upon by many forces which give the body an acceleration that is usually much larger than $a_{0}$.
 
Because of that, it was considered very close to impossible. But recently it was found that some specially arranged conditions allows us to test  the hypothesis in a normal terrestrial laboratory.
 
 To do this, we have to provide such conditions that would lead to cancellation of all large forces. All those large forces that act on our test body should cancel almost exactly leaving only a tiny residue. This is the main idea.
 
To do that is of course not easy. Basically, it could be explained as follows. When we are on the Earth, each body is acted upon by the inertial forces due to the  the Earth's spinning around its axis, orbiting around the Sun (centrifugal forces) etc. So because the Earth motions, we have several inertial forces. And the main problem is, how to make these  forces cancel?
 
This is exactly what we are interested in. It appears that the answer depends crucially on the state of motion of the test body. Depending on how the body moves, the conditions of ``entry'' into the MOND regime could be very different. And the corresponding experiments could vary   wildly  in their difficulty. This is still a very much open area.  Initially, it was important to focus on the ``proof-in-principle'' that such experiments are at all possible, laying aside more practical considerations of the required costs and efforts.   Because of this, the attention was drawn to the {\em conceptually} the simplest case when  the test body is at rest (relative to the Earth). We emphasized from the beginning that this is {\em only one of the many possibilities} \cite{Ig1}. It may or may not be practical, but if not, there are plenty of other options to contemplate. At least three of them have been discussed in the literature \cite{Ig1, DFP}. The main thrust of this paper is to draw attention to these alternatives. However, at  present these alternative options are largely in the nascent stage and many details are yet to be filled in. (The reason is that we are talking here of a challenging problem requiring collective efforts of both theorists and experimenters.)

Let us start with reviewing briefly the ``rest'' case---just to set the stage and for comparison with other scenarios.

If we ask if the inertial forces can cancel in the static case, then the net result is that this cancellation can be realized but not always and not everywhere. Only at a certain time and at a certain place on the Earth the conditions could satisfy the cancellation requirement. That looks like very lucky circumstances, they are very rare and special occasions.
 
These places lie on the latitude about 80 degrees to the north or to the south---not the places where you would normally go for a holiday!
But such places are not new for physicists: in the 90Õs they drilled a borehole in Greenland to see if there are deviations from the gravitation law (the so-called ``fifth forceÕÕ).
 
However, we may prefer to perform the experiment in the comfort of our laboratory at home. Then, we still have to exploit the same idea of cancellation of forces, but in this case different forces will be involved. This leads to the so-called ÒCCC setupÓ   where CCC symbolizes the Cancellation between the Coriolis and Centrifugal forces \cite{Ig1}.
 
In this case, there is no restriction on the laboratory location. But the restriction on time (only twice a year) remains. Technically, this stems from the fact that the centrifugal forces are due to the Earth's spin.
 
Finally, there are at least two approaches to doing the test at any time and any place. One is to move a test body along a special trajectory with a prescribed speed and acceleration \cite{Ig1}.
 
Another approach \cite{DFP} exploits the same basic idea of cancellation, but in a different experimental set-up. The key point is to introduce extra, man-made centrifugal forces into play (in addition to the usual, terrestrial ones). To realize this, a spinning object, such as a ring is needed. If the rotation of the ring is carefully  controlled, then there is a chance to achieve the desired cancellation at any time and any place.
 
 
All four setups proposed so far are technically challenging and it is hard to predict the ultimate  winner. But the game is exciting: itÕs not every day that a 300-year old law can be challenged!
\section{Which reference frame?}
It is extremely important to clarify \footnote{Overlooking this point  leads to a misguided perception that MOND has been already ruled out experimentally.} which reference frame should be used for designing  the MOND tests \cite{Ig1,Ig2}. The main point is that the usual lab frame, i.e., the frame fixed o the Earth is not a suitable one. Although such a frame can be treated as inertial for many purposes,  it is not inertial when we talk about MOND because of the smallness of the key acceleration scale $a_0$. Compared to that scale, the usually insignificant inertial forces  become huge. So even if we manage to arrange for a test body that has a very small acceleration relative to the lab frame, and look for possible deviations from Newtonian dynamics, this would be an interesting experiment, but it would not be possible to confirm or refute MOND using its outcome (cf. \cite{AV, G, GE, SS, RS}).

What is needed, is the Galactic reference frame rather than the laboratory one. Are the alternative choices possible? For example, what about  a frame with the origin at the center of mass of the Local Group of galaxies? Fortunately, the acceleration due to the neighbour galaxies is much smaller than $a_0$. For example, the acceleration due to the Andromeda galaxy is less than $10^{-12}\; m\, s^{-2}$. Therefore, although such an alternative is possible, it would not affect the results very much.

The above intuitive argument can be formalized as follows \cite{Ig2}.

Suppose we have two frames of reference: S (inertial) and $S'$ (non-inertial). Let $S'$ moves with acceleration ${\bf b}$ with respect to S. The equation of motion in S reads:
\begin{equation}
\label{a1}
\mathbf{F}=m\mathbf{a}\mu(a/a_0).
\end{equation}
Here $\bf{a}$ is the test body acceleration relative to S.

  Assuming that $a_0$ is invariant, in the S$' $ frame
we have the following  equation of motion:
 \begin{equation}
\label{a2}
\mathbf{F}=m\mathbf{a'}\mu(a'/a_0)+m{\bf b},
\end{equation}
where ${\bf a'}={\bf a}-{\bf b}$ represents the test body acceleration   relative  to  S$'$.
It is easily seen that the two equations (\ref{a1}) and (\ref{a2}) cannot be satisfied simultaneously for all $\bf{a}$ and $\bf{b}$. Indeed, if ${\bf a}=0$ then 
\begin{equation}
\label{}
m\mu(b/a_0)=m
\end{equation}
for all $\bf{b}$ which means that $\mu(z)=1$ for all $z$. This contradiction proves that it is not possible to assume that the value of the critical acceleration is independent of the reference frame. In other words, modified Newton's law should be formulated relative  to the Galactic reference frame, and not  relative to the laboratory reference frame.


 \section{The SHLEM effect}
 To make the paper self-consistent, we briefly recall the essence of what was dubbed ``the SHLEM effect'' in \cite{Ig1}.  This effect refers to the possibility of testing MOND using a {\em static} probe.  The essence of the effect is that the cancellation equations yield the solutions that are strictly localized, both in time and in space.
 
 Time-wise, the solutions occur around the equinoxes, and space-wise---in the vicinity of $80^o$ latitudes.
 
 The effect itself consists in a spontaneous, tiny displacement of the test body at those special instants of  time and at those special points on Earth.
 
 In addition to the ground-based experiments, the approach proposed in \cite{Ig1} allows one to discuss a possibility of using an Earth's satellite for testing MOND.
 This would require a very high altitude orbit $R_{orbit}\simeq R_{Earth}(g/a_s)^{1/2}\simeq40R_{Earth}$ and the inclination $\simeq 23^o27'$ (so that the orbit is in the ecliptic plane). For such a satellite, entering the MOND regime and, thus, the violation of Newton's 2nd law would be expected to occur once per revolution around the Earth, i.e., every 15 days.  Further study is needed to understand whether the anomalies in the satellite's motion due to this effect would be observable or not. In particular, lunar and planetary as well as non-gravitational effects (see, e.g., \cite{MNF, Ior} should be taken into account.

  \section{Beyond SHLEM }
  It is important to realize that using the  SHLEM set-up is certainly not a unique way to test MOND.  Several alternatives were discussed, along with SHLEM, already in Ref. \cite{Ig1}.  Another scenario was proposed in Ref. \cite{DFP}. Even this list is unlikely to be exhaustive, and new opportunities could well be contemplated. However, the problem is challenging, and collaboration between experimentalists and theorists would probably  be needed to turn the general ideas into specific proposals and to give it a significant boost.
  
  The necessary and sufficient condition for entering the MOND regime in the lab reads \cite{Ig1}:
\begin{equation}
\label{2}
\mathbf{a}_{lab}\approx -\mathbf{a}_1(t)-\bm{\om}\times(\bm{\om}\times(\mathbf{r}+\mathbf{r}_1))-2 \bm{\om}\times\mathbf{v}-\mathbf{a}_2.
\end{equation} 
  Here, $\mathbf{a}_1$ is the acceleration of the Earth's centre relative to the heliocentric reference frame, $\bm{\om}$ is the the Earth's angular velocity,   $\mathbf{a}_2$ is the Sun's acceleration with respect to $S_0$, $\mathbf{r},\; \mathbf{v}=\dot{\mathbf{r}}$, and $\mathbf{a}_{lab}=\ddot{\mathbf{r}}$ are the position, velocity, and acceleration of the test body relative to the lab  frame;  $\mathbf{r}_1$ is the position vector of the origin of the lab frame relative to the terrestrial frame with the origin at the Earth's centre.
  
 The general solution of the above equation requires knowledge of the functions $\mathbf{a}_1(t)$ and $\mathbf{a}_2(t)$, which can be obtained from astronomical data. Once they are given, the equation should be solved numerically with the required accuracy.  To get an idea of such a solution, let us make these modelling assumptions:
  
   (1) Acceleration $\mathbf{a}_2(t)$  is neglected (in other words, we assume that the heliocentric frame is  inertial). (2) Acceleration $\mathbf{a}_1(t)$ is taken as a harmonic oscillation with the frequency $\om_1 = 2\pi /(1\; yr)$ (i.e., the Earth orbit's eccentricity and the lunar effects  are ignored). (3) We assume that the direction of $\bm{\om}$ (taken as $z$-axis) is  perpendicular to the ecliptic. 
   
   Under the above assumptions, the  general solution of Eq.~(\ref{2}) becomes
\begin{equation}
\label{9}
\ddot x \approx  (x_1+x_0) \om^2+2 v_{0y}\om - R\om_1^2 + y_1\om^3t - 3v_{0x}\om^3 t - x_1 \om^4 t^2,
\end{equation}
\begin{eqnarray*}
\ddot y \approx & (y_1+y_0) \om^2 - 2 v_{0x}\om  - 2 x_1\om^3t - 3 v_{0y}\om^2 t +\\ & v_{0x}\om^2 t + 3R \om_1^2\om t - y_1\om^4 t^2,
\end{eqnarray*}
where   
$x_0$, $y_0$, $v_{0x}$, $v_{0y}$ are the initial position and velocity of the test body; $x_1, \; y_1$ are the coordinates of the origin of the lab frame relative to the Earth's centre, and  $R$ is the Earth-Sun distance.

Using integration once or twice, Eq.~(\ref{9}) will yield  the trajectory and the velocity  of the test body which will satisfy the conditions of MOND regime {\em at all times} and not just at the special instants around the equinoxes. Also, the restrictions on the laboratory location disappear. An experiment based on the model Eq.~(\ref{9})could be performed {\em anywhere}.  Thus we have a strong argument  that space-time unrestricted  set-ups could be possible. However, a lot of further work would be required to find out if an actual experiment could be designed along these lines.

Schematically, the set-up could be as follows. Take a test body, and move it precisely along the ``MOND trajectory'' with the precisely ``MOND velocity''. Then the MOND prediction is that anomalous behaviour (e.g., unaccountable residues in the position or velocity data) would be observed.

The next category of MOND tests could be termed ``space-unrestricted, time-restricted'' experiments. Like the above set-up, they could be performed anywhere on Earth, but only at certain specific times around the equinoxes. An example in this category could be the CCC set-up \cite{Ig1} based on the cancellation between the centrifugal and Coriolis inertial forces.

As one of the specific realizations of the CCC scenario, we may consider 
testing   MOND in a time-of-flight laboratory experiment. Again, we would like to stress  that the TOF tests are just one of the many possible examples of the  ``space-unrestricted, time-restricted'' category.

For this experiment to succeed,  many factors should be taken into account, some of which are in conflict with each other. Here, we restrict ourselves to just  listing all these factors leaving further details for future work:

Velocity magnitude

Velocity direction

Accuracy of velocity control

Baseline length

Accuracy of time measurement

Projectile type


Statistics

Gravity compensation accuracy


Gravity compensation mechanism

Vacuum

Cryogenics

Screening of external fields


Relative or absolute measurement?

\section{Outside MOND regime?}
The above mentioned scenarios are very different and require entirely separate set-ups. Yet even they do not exhaust all possible routes to testing   MOND in the laboratory.  All the above options, although differing in their approach, shared one common thing: they attempted to create conditions for entry into MOND regime, so we could probably call all of them collectively the ``inside MOND'' experiments.

However, seeing how difficult it is to get inside, we could also try to test MOND using high-precision measurements {\em without entering the MOND regime}, but being, in some sense, close to it. These attempts could be termed ``outside MOND'' tests.

It remains to be seen whether a competitive set-up can be designed along these lines .

 \section{Novel set-up}Due to a recent breakthrough in
 precision accelerometry,  
 a completely new set-up can be imagined. A new accelerometer built at NIST has the resolution of $10^{-18} \;\rm{m/\sqrt{Hz}}$ over the frequency range of several kHz \cite{Cer}. On the other hand,  characteristic displacements of a test body due to the SHLEM effect are of the order of $10^{-14} \;\rm{m}$ \cite{Ig2}. Therefore, using accelerometers could be a promising direction to pursue in designing a sensitive SHLEM experiment.
 \section{Physics and Astrophysics}
The MOND theory is so unusual for many physicists that the astrophysical evidence alone, no matter how strong, would probably be not sufficient to convince the sceptics.  What is missing is the evidence from the lab. The experimental proof of MOND will be hard to deny or ignore.

In fact, a parallel with the dark matter case is entirely appropriate here. Searches for DM are conducted on Earth as vigorously as in cosmos.  MOND should be given the same chance.

The main thing is that searches for MOND effects, while challenging, are not prohibitively hard. But they do require a lot of collaboration between experimenters and theorists. It seems that only team efforts can bring further progress here.

       \end{document}